\documentclass{ws-procs9x6}

\begin{document}

\title{RESULTS FROM THE NEMO~3 EXPERIMENT}

\author{L. V\'{A}LA\footnote{on behalf of the NEMO Collaboration}}

\address{IEAP, Czech Technical University in Prague,\\
Horsk\'a 3a/22, Prague, CZ -- 128~00, Czech Republic\\
E-mail: ladislav.vala@utef.cvut.cz}

%

\begin{abstract}
The aim of the NEMO~3 experiment 
is the search for neutrinoless double beta decay 
and precise measurement of two-neutrino double beta decay
of several isotopes.
The experiment has been taking data since 2003.
Since no evidence for neutrinoless double beta decay
of $^{100}$Mo and $^{82}$Se has been found,
a 90\% C.L. lower limit on the half-life of this process
and corresponding upper limit 
on the effective Majorana neutrino mass
are derived.
The data are also interpreted in terms of alternative models,
such as weak right-handed currents or Majoron emission.
In addition,
NEMO~3 has performed precision measurements of
the two-neutrino double beta decay for seven different isotopes.
The most recent experimental results of NEMO~3 are presented in this paper.
\end{abstract}


\bodymatter

\section{Introduction}
The neutrino oscillation experiments demonstrated in the last decade
convincing evidence for neutrino oscillations
proving the finite neutrino masses and mixing.
However,
these experiments are only sensitive to the difference
in the square of the neutrino masses,
\( \Delta m^2_{ij} = | m^2_i - m^2_j | \),
therefore they do not provide information 
on the absolute scale of the neutrino masses
and are not sensitive to the nature of neutrinos
Nevertheless,
the detection and study of neutrinoless double beta decay 
is the only way able to answer
the following fundamental questions:
(i) neutrino nature (Majorana or Dirac particle?),
(ii) absolute neutrino mass scale,
(iii) type of neutrino mass hierarchy (degenerated, inverted, or normal?),
(iv) CP violation in the lepton sector.

Double beta decay ($\beta\beta$) is a transition 
from nucleus $(A,Z)$ to $(A,Z+2)$
and it can occur in different channels:
(a) two-neutrino double beta decay ($2\nu\beta\beta$) 
with emission of two $e^-$ and two $\bar{\nu_e}$,
(b) neutrinoless double beta decay ($0\nu\beta\beta$)
with emission of two $e^-$ and 
(c) neutrinoless double beta decay with Majoron emission
($0\nu\chi\beta\beta$)
with two $e^-$ and Majoron $\chi$ 
(massless Goldstone boson)
in the final state.
The mode (a) is a process of second order
allowed in the Standard Model (SM) 
which has been observed for several nuclei.
Nevertheless,
processes (b) and (c) are violating lepton number conservation 
by two units and involve new physics beyond the SM.

The NEMO~3 experiment
(NEMO = Neutrino Ettore Majorana Observatory)
is devoted to the search for $0\nu\beta\beta$ decay
and to the accurate measurement of $2\nu\beta\beta$ decay.
For this goal,
the experiment combines two detection techniques:
calorimetry and tracking.
Such approach allows us at the same time
unambiguous identification
of $e^-$, $e^+$, $\gamma$, and $\alpha$-particles
provided by a wire tracking chamber
and energy and time measurements of particles 
with a calorimeter.

\section{NEMO~3 detector}

The NEMO~3 detector [\refcite{ARN05a}]
is installed and currently running 
in the Fr\'ejus Underground Laboratory (4800~m w.e.) in France.
The experimental set-up is cylindrical in design,
is divided into twenty equal sectors and 
with $\gamma$ and neutron shielding
it has about 6~m in diameter and 4~m in height.

The tracking wire chamber is made of 6180 open octagonal drift cells
operating in Geiger mode (Geiger cells).
It is filled  with a gas mixture of 
95\% He, 4\% ethyl-alcohol and 1\% Ar.
The Geiger cells provide a three-dimensional measurement of 
the charged particle tracks 
by recording the drift time and the two plasma propagation times.

The calorimeter, 
which surrounds the wire chamber,
is composed of 1940 plastic scintillators coupled by light-guides 
to very low-radioactivity PMTs.
The energy resolution $\sigma_E/E$ of the calorimeter
ranges from 6.0\% to 7.5\% for 1~MeV electrons,
while the time resolution is 250~ps.

Seventeen sectors of NEMO~3 accommodate almost 10~kg of 
the following, highly enriched (95\% -- 99\%) 
$\beta\beta$ decay isotopes:
$^{100}${Mo} (6914~g), $^{82}${Se} (932~g), 
$^{116}${Cd} (405~g), $^{130}${Te} (454~g), 
$^{150}${Nd} (34~g), $^{96}${Zr} (9~g), and $^{48}${Ca} (7~g).
In addition, 
three sectors are also used for
external background measurement
and are equipped with pure Cu and natural Te.
All these isotopes are produced
in the form of thin foils
and are placed in the central vertical plane of each sector.

For the $e^- / e^+$ recognition,
the detector is surrounded by a solenoidal coil 
which generates a vertical magnetic field of 25~Gauss.
Moreover,
NEMO~3 is covered by two types of shielding 
against external $\gamma$-rays and neutrons.
Finally,
the whole experimental set-up is closed inside a ``tent'', 
which is supplied with radon-free air from a radon trapping facility.
Radon is trapped and then decays inside a tank
filled with 1~ton of charcoal
cooled down to $-50^{\circ}$C.
This facility decreases the radon level of 
the air from the laboratory (15 -- 20~Bq/m$^3$) 
by a factor of $\sim 1000$.
The radon trapping facility has been operating 
in the laboratory since October 2004.

\section{Measurement of $\beta\beta$ decay and backgrounds}
A candidate for a $\beta\beta$ decay is a two-electron event
which is defined with the following criteria:
two tracks coming from the same vertex in the source foils,
the curvature of the tracks corresponds to a negative charge,
each track has to be associated with a fired scintillator,
and the time-of-flight has to correspond to 
the case of two electrons emitted at the same time 
from the same vertex.
In addition
a threshold of 200~keV is applied on energy of each electron.
Finally,
it is also required 
that there is no delayed Geiger cell hit
close to the event vertex
in order to suppress background from 
$^{214}$Bi decay inside the tracking detector.

The energy window of interest for the $0\nu\beta\beta$ decay for
both $^{100}$Mo and $^{82}$Se is set to $(2.8 - 3.2)$~MeV
and the complete study of background contribution
in this window has been performed. 
The level of each background component has been directly measured
from data using different analysis channels.
The dominant background during the first running period
from February 2003 to September 2004, Phase~I,
was due to radon diffusion into the tracking wire chamber.
Nevertheless,
during 
Phase~II (since November 2004 up to now),
the radon level inside NEMO~3
has been reduced by a factor of ten
thanks to the radon trapping facility.
Remaining low radon activity inside NEMO~3 
is due to degasing of detector components.

\section{NEMO~3 results}
\subsection{$2\nu\beta\beta$ decay}
The $2\nu\beta\beta$ decay of $^{100}$Mo and $^{82}$Se
is measured with high acuracy with NEMO~3.
The obtained half-lives for Phase~I data (389~d) are
\( T_{1/2} = [ 7.11 \pm 0.02(stat) \pm 0.54(syst)] \times 10^{18} \)~y
for $^{100}$Mo and
\( T_{1/2} = [ 9.6 \pm 0.3(stat) \pm 1.0(syst)] \times 10^{19} \)~y
for $^{82}$Se [\refcite{ARN05b}].
The half-lives for the other five $\beta\beta$ isotopes
have been also derived from data 
and are summarised in Tab.~\ref{tab:results-2nubb}.
Measurements of this process are important for 
nuclear theory as they allows us to
reduce the uncertainties on the nuclear matrix elements (NME).

\begin{table}[htb]
  \tbl{Measured half-lives of $2\nu\beta\beta$ decay and $S/B$ ratio.}
  {\begin{tabular}{rllr@{}l}
    \toprule
    \multicolumn{2}{l}{Isotope and transition} & $T_{1/2}$  & \multicolumn{2}{l}{$S/B$}\\
    \colrule
    $^{100}$Mo & (g.s. $\rightarrow$ g.s.) 
                 & $[7.11 \pm 0.02(stat) \pm 0.54(syst)] \times 10^{18}$~y & 40 & \\
               & (g.s. $\rightarrow 0^+_1$)
                 & $[5.7^{+1.3}_{-0.9}(stat)\pm 0.8(syst)]\times 10^{20}$~y & 3 & .0 \\
               & (g.s. $\rightarrow 2^+_1$)
                 & $> 1.1 \times 10^{21}$~y at 90\% C.L.                    &   & \\
    $^{82}$Se  & (g.s. $\rightarrow$ g.s.) 
               & $[9.6 \pm 0.3(stat) \pm 1.0(syst)] \times 10^{19} $~y    & 4 & .0 \\
    $^{116}$Cd & (g.s. $\rightarrow$ g.s.)
               & $[2.8 \pm 0.1(stat) \pm 0.3(syst)] \times 10^{19} $~y    & 7 & .6 \\
    $^{150}$Nd & (g.s. $\rightarrow$ g.s.)
               & $[9.7 \pm 0.7(stat) \pm 1.0(syst)] \times 10^{18} $~y    & 2 & .4 \\
    $^{96}$Zr  & (g.s. $\rightarrow$ g.s.)
               & $[2.0 \pm 0.3(stat) \pm 0.2(syst)] \times 10^{19} $~y    & 0 & .9 \\
    $^{48}$Ca  & (g.s. $\rightarrow$ g.s.)
               & $[3.9 \pm 0.7(stat) \pm 0.6(syst)] \times 10^{19} $~y    & 15& .7 \\
    $^{130}$Te & (g.s. $\rightarrow$ g.s.)
               & $[7.6 \pm 1.5(stat) \pm 0.8(syst)] \times 10^{20} $~y    & 0 & .25 \\
    \botrule
  \end{tabular} }
  \label{tab:results-2nubb}
 \end{table}

\subsection{$0\nu\beta\beta$ decay}
In the case of $^{100}$Mo,
there are 14 events observed in the energy window of interest
while 13.4 events were expected from backgrounds
for combined Phase~I and II data (693~d).
The situation is similar for $^{82}$Se as 7 events are observed
and 6.2 expected for the same period.
Thus,
resulting half-life limits at 90\% C.L. are 
\( T_{1/2} > 5.8 \times 10^{23} \)~y for $^{100}$Mo
and \( T_{1/2} > 2.1 \times 10^{23} \)~y for $^{82}$Se.
If using the NME from Refs. [\refcite{KOR07a,KOR07b,ROD07}],
the following limits on the effective neutrino mass are derived:
\( \langle m_{\nu} \rangle <(0.6 - 0.9) \)~eV for $^{100}$Mo
and \( \langle m_{\nu} \rangle <(1.2 - 2.5) \)~eV for $^{82}$Se.
We determined also limits for alternative models
assuming weak right-handed currents (V+A) and 
the $0\nu\chi\beta\beta$ decay channel 
[\refcite{ARN06}].
All these results are given in Tab.~\ref{tab:results-0nubb}.

\begin{table}[htb]
  \tbl{Limits at 90\% C.L. for different $0\nu\beta\beta$ decay modes 
    for $^{100}$Mo and $^{82}$Se.}
  {\begin{tabular}{lll}
    \toprule
    Decay mode & $^{100}$Mo & $^{82}$Se \\
    \colrule
    $0\nu\beta\beta \,\rm{(V - A)} \, (\rm{g.s.} \rightarrow \rm{g.s})$ 
          & $T_{1/2} > 5.8 \times 10^{23}$~y 
          & $T_{1/2} > 2.1 \times 10^{23}$~y \\
          & $\langle m_{\nu} \rangle <(0.6 - 0.9)$~eV 
          & $\langle m_{\nu} \rangle <(1.2 - 2.5)$~eV \\
          & $\langle m_{\nu} \rangle <(2.1 - 2.7)$~eV 
          & $\langle m_{\nu} \rangle <(2.6 - 3.2)$~eV \\
   $0\nu\beta\beta \,\rm{(V - A)} \, (\rm{g.s.} \rightarrow 0^+_1)$ 
          & $T_{1/2} > 8.9 \times 10^{22}$~y  &  \\
   $0\nu\beta\beta \,\rm{(V - A)} \, (\rm{g.s.} \rightarrow 2^+_1)$ 
          & $T_{1/2} > 1.6 \times 10^{23}$~y  &  \\
    $0\nu\beta\beta \, \rm{(V + A)} \, (\rm{g.s.} \rightarrow \rm{g.s})$ 
          & $T_{1/2} > 3.2 \times 10^{23}$~y 
          & $T_{1/2} > 1.2 \times 10^{23}$~y \\
    $0\nu\chi\beta\beta \, (n = 1)$ 
          & $T_{1/2} > 2.7 \times 10^{22}$~y 
          & $T_{1/2} > 1.5 \times 10^{22}$~y  \\
    $0\nu\chi\beta\beta \, (n = 2)$ 
          & $T_{1/2} > 1.7 \times 10^{22}$~y
          & $T_{1/2} > 6.0 \times 10^{21}$~y  \\          
    $0\nu\chi\beta\beta \, (n = 3)$ 
          & $T_{1/2} > 1.0 \times 10^{22}$~y 
          & $T_{1/2} > 3.1 \times 10^{21}$~y  \\
    $0\nu\chi\beta\beta \, (n = 7)$ 
          & $T_{1/2} > 7.0 \times 10^{19}$~y 
          & $T_{1/2} > 5.0 \times 10^{20}$~y  \\
    \botrule
  \end{tabular} }
  \label{tab:results-0nubb}
 \end{table}

However, these results date back to 2006
because the NEMO Collaboration decided 
to perform blind analysis with mock data.
We plan to open the box 
and update the results by the summer of 2008
and once again after the end of the experiment by 2010.

\subsection{Double beta decay of $^{100}$Mo to excited states}
The $\beta\beta$ decay of $^{100}$Mo
to excited $0^+_1$ and $2^+_1$ states of $^{100}$Ru 
has been also studied with the NEMO~3 detector.
The obtained half-life for the $2\nu\beta\beta$ decay
to the $0^+_1$ state is
\( T_{1/2} = [5.7^{+1.3}_{-0.9}(stat)\pm 0.8(syst)]\times 10^{20} \)~y
[\refcite{ARN07}].
This value is in a good agreement with previous measurements
[\refcite{BAR95,BRA01}].
In addition,
the $T_{1/2}$ limits have been determined for 
the $0\nu\beta\beta$ and $2\nu\beta\beta$ decay to the $2^+_1$ state 
and for the $0\nu\beta\beta$ decay to the $0^+_1$ state
(see Tabs.~\ref{tab:results-2nubb} and \ref{tab:results-0nubb}).

\section{Conclusions}
The NEMO~3 detector has been routinely taking data since February 2003.
The $2\nu\beta\beta$ decays of $^{100}$Mo and $^{82}$Se 
have been measured with very high statistics 
and better precision than in previous experiments.
The $2\nu\beta\beta$ 
half-lives have been obtained also for
$^{116}${Cd}, $^{130}${Te}, $^{150}${Nd}, $^{96}${Zr}, and $^{48}${Ca}.

No evidence for $0\nu\beta\beta$ decay has been found in
combined data for Phase~I and II corresponding to 693 days.
The $T_{1/2}$ limits at 90\% C.L.
for different modes of this decay are summarised 
in Tab.~\ref{tab:results-0nubb}.
We obtained currently the best limits for 
the $0\nu\chi\beta\beta$ decay of $^{100}$Mo and $^{82}$Se

At present time,
the analysis of the Phase~II data without radon is in progress
and improved $2\nu\beta\beta$ and $0\nu\beta\beta$ results
will be published in 2008.
The expected half-life limits for the $0\nu\beta\beta$ decay
by the end of the experiment in 2010 are 
$ T_{1/2} > 2 \times 10^{24}$~y for $^{100}$Mo and 
$ > 8 \times 10^{23}$~y for $^{82}$Se.

%
\section*{Acknowledgment}
A portion of this work has been supported by 
the Grant Agency of the Czech Republic
(grant No.~202/05/P293).

%

\end{document}